\documentclass[3p,times,twocolumn]{elsarticle}

\usepackage{ecrc}

\volume{00}

\firstpage{1}

\journalname{Nuclear Physics B Proceedings Supplement}

\runauth{Andr\'e H.~Hoang et al.}

\jid{nuphbp}

\jnltitlelogo{Nuclear Physics B Proceedings Supplement}

\usepackage{amssymb}
\usepackage{amsthm}
\usepackage{amsmath}
\usepackage{amsfonts}
\usepackage[small]{subfigure}

\newcommand{\df}{\mathrm{d}}
\newcommand{\nn}{\nonumber}

\usepackage[figuresright]{rotating}

\begin{document}

\begin{frontmatter}



\dochead{}

\title{State-of-the-Art Predictions for C-parameter and a Determination of $\alpha_s$}


\author{Andr\'e H.~Hoang$^{1,2}$, Daniel W.~Kolodrubetz$^3$, Vicent Mateu$^{1,*}$, Iain W.~Stewart$^3$}
\address{$^1$University of Vienna, Faculty of Physics, Boltzmanngasse 5, A-1090 Wien, Austria}\vspace{0.1cm}
\address{$^2$Erwin Schr\"odinger International Institute for Mathematical
Physics, University of Vienna, Boltzmanngasse 9, A-1090 Vienna, Austria}
\address{$^3$Center for Theoretical Physics, Massachusetts Institute of
  Technology, Cambridge, MA 02139, USA\vspace{0.1cm}\\
\flushleft $^*$ Speaker\\[1pt]

Preprint numbers: UWTHPH 2014-24, MIT--CTP 4601, LPN14-122}
\begin{abstract}
The C-parameter event-shape distribution for $e^+\, e^-$ annihilation into hadrons is computed in the framework of SCET including input from fixed-order perturbation theory. We calculate all missing ingredients for achieving N${}^3$LL resummation accuracy in the cross section, which is then matched onto $\mathcal{O}(\alpha_s^3)$ fixed-order results. Hadronization power corrections are incorporated as a convolution with a nonperturbative shape function. Wide-angle soft radiation effects introduce an $\mathcal{O}(\Lambda_{\rm QCD})$ renormalon ambiguity in the cross section, which we cure by switching to the Rgap short-distance scheme. We also include hadron mass effects, but find their effect is rather small. Performing fits to the tail of the C-parameter distribution for many center of mass energies we find that the strong coupling constant is \mbox{$\alpha_s(m_Z)=0.1123\pm 0.0015$}, with $\chi^2/\rm{dof}=0.99$.
\end{abstract}


\end{frontmatter}


\section{Introduction}
\label{sec:introduct}
The LEP $e^+\, e^-$ collider, previously located at CERN, has delivered an enormous amount of highly accurate experimental data, which can be used to explore the theory of strong interactions in its high-energy regime. To study Quantum Chromodynamics (or QCD) at high energies one needs to deal with jets: highly boosted and collimated bunches of particles that can be seen as the remnants of the underlying partons created at very short distances. One appealing strategy for describing jet dynamics is through event shapes, infrared- and collinear-safe observables which are constructed from the energy and momenta of all the produced hadrons (in this sense event shapes are global quantities). They are designed to measure  geometrical properties of the distribution of particles, and in particular they quantify how ``jetty'' the final state is. Additionally, being global observables, it is possible to compute high-order perturbative corrections, carry out log resummation to higher order, show factorization and exponentiation properties, and use factorization to control power corrections.

One of the main uses of event-shape distributions is the determination of the strong coupling constant $\alpha_s$. The
advantage of event shapes over other inclusive observables is that they are essentially proportional to $\alpha_s$, rather than probing $\alpha_s$ via corrections to a leading term (as is the case, for example, of DIS or the total hadronic cross section). Thus, event shapes are very sensitive to the strong coupling constant. On the other hand, event shapes are afflicted by nonperturbative power corrections and by large double Sudakov logarithms, which necessitate resummation.

Here we study the C-parameter which can be written as~\cite{Parisi:1978eg,Donoghue:1979vi}:
\begin{equation} \label{eq:Cdef}
C=\frac{3}{2} \, \frac{\sum_{i,j} | \vec{p}_i | | \vec{p}_j | \sin^2 \theta_{ij}}
{\left( \sum_i | \vec{p}_i | \right)^2}\,.
\end{equation}
It is interesting to compare C-parameter with thrust~\cite{Farhi:1977sg}, 
\begin{equation}
\tau \,=\, 1-T \,=\,
\min_{\vec{n}} \left( 1 -\, \frac{ \sum_i|\vec{n} \cdot \vec{p}_i|}{\sum_j |\vec{p}_j|} \right)\,,
\end{equation}
where $\vec{n}$ is referred to as the thrust axis. The three main differences between $C$ and $\tau$ are:
a)~$C$ does not require a minimization  procedure for its computation, whereas $\tau$ does (namely finding the thrust axis, event by event)\,; b) $C$ is defined through a double sum, whereas $\tau$ sums only particle by particle\,; c)~the fixed-order prediction of C-parameter develops an integrable singularity at $C_{\rm shoulder}=0.75$, whereas thrust is always smooth. By shoulder we refer to the fact that the partonic  cross section attains an integrable singularity~\cite{Catani:1997xc} at $C_{\rm shoulder}$, and only non-planar configurations contribute for $C>C_{\rm shoulder}$.\,\footnote{In Ref.~\cite{Catani:1997xc} it is shown that soft gluon resummation at $C_{\rm shoulder}$ makes up for a smooth distribution at LL order.} There are also a number of similarities between $C$ and $\tau$, and perhaps the most remarkable one is that in the dijet limit ($C,\tau\ll 1$) and up to and including NLL resummation, both partonic cross sections are related in a simple way, which can be schematically expressed as follows:
\mbox{$\tau_{\rm NLL}=C_{\rm NLL}/6$}~\cite{Catani:1998sf}. Some other similarities will be highlighted later.

Previous analyses of the thrust distribution using SCET at N$^3$LL and analytic power corrections have found rather
small (albeit precise) values of $\alpha_s$~\cite{Abbate:2010xh,Abbate:2012jh}\,\footnote{Other lower-order resummation
event-shape analyses have also found small (although less accurate) values of $\alpha_s$ \cite{Gehrmann:2009eh,
Gehrmann:2012sc}.}. Two motivations for carrying out this new analysis are providing an additional determination
of $\alpha_s$ and studying the universality of the leading power correction between thrust and C-parameter.
In this proceedings we summarize work done in Refs.~\cite{Hoang:2014wka,Hoang:2015hka}.

\section{Theoretical developments}
\label{sec:theoretical}
Until a few years ago, theoretical uncertainties were larger than the corresponding experimental ones and hadronic power
corrections were not understood from ab-initio QCD considerations. The situation on the theory side
has dramatically changed with the advent of Soft-Collinear Effective Theory (SCET)~\cite{Bauer:2000ew, Bauer:2000yr,
Bauer:2001ct, Bauer:2001yt}. This effective field theory separates the relevant physics occurring at the various scales
which play a role when jets are being produced: hard scale $\mu_H$, of the order of the center of mass energy $Q$
(describes the production of partons at very short distances), jet scale $\mu_J\sim Q\sqrt{C/6}$ (describes the formation
and evolution of jets at intermediate energies), and the soft scale $\mu_S\sim QC/6$ (describes wide angle soft radiation
at longer distances). All three scales are widely separated for $C\ll 1$, and there is one function associated to each one of
them: the hard coefficient $H_{\!Q}$ (the modulus square of the QCD to SCET matching coefficient), which is common
to all event-shape factorization theorems\,; the Jet function $J_\tau$ (built up with collinear Wilson lines), which is common
for thrust, C-parameter~\cite{Hoang:2014wka} and Heavy Jet Mass ($\rho$)~\cite{Chien:2010kc}\,;
and the Soft function $S_{\!C}$ (defined through soft
Wilson lines), which in general depends on the specific form of the event shape. Whereas the former
two are perturbative ($\mu_H, \mu_J \gg \Lambda_{\rm QCD}$), permitting the calculation of the hard and jet functions in powers of $\alpha_s$, the soft function also has nonperturbative corrections that need to be accounted for ($\mu_S\gtrsim \Lambda_{\rm QCD}$). Renormalization evolution among the three scales sums up large logarithms to all orders in perturbation theory. It turns out that the anomalous dimensions for the soft function for $C$ and $\tau$ are identical, see~\cite{Hoang:2014wka}.

The soft function can be further factorized into a partonic soft function $\hat S_{\!\widetilde C}$, calculable in perturbation
theory, and a nonperturbative shape function $F_{\!C}$, which has to be obtained from fits to data\,\footnote{Power corrections
for the C-parameter distributions have been studied in other frameworks, see e.g.\
Refs.~\cite{Gardi:2003iv,Korchemsky:2000kp,Dokshitzer:1995zt}.}. The treatment of hadronic
power corrections greatly simplifies in the tail of the distribution, defined by $QC/6\gg \Lambda_{\rm QCD}$, where the
shape function can be expanded in an OPE. The leading power correction is parametrized by $\Omega_1^C$, the first moment of the shape function. Interestingly, if one ignores hadron mass
effects~\cite{Salam:2001bd,Mateu:2012nk}, this matrix element can be related to the corresponding one in thrust in a trivial
manner: \mbox{$\Omega_1^\tau/2 = \Omega_1^C/(3\pi)\equiv \Omega_1$}~\cite{Lee:2006nr}. The main effect of this leading power correction is a shift of the cross section, \mbox{${\df\hat\sigma} (C) \to {\df\hat\sigma}(C \,-\, \Omega_1^C/Q)$}. When presenting the results of our fits, we will employ the power correction parameter $\Omega_1$ to ease comparison.

The leading SCET factorization for the partonic \mbox{C-parameter} distribution can be written
as~\cite{Bauer:2008dt,Hoang:2015hka}:
\begin{align}\label{eq:fact-singular}
\!\!\!\!\!
\frac{1}{\sigma_0}\frac{\df \hat\sigma_{\rm s}}{\df C}=\frac{Q}{6} H_{\!Q}(Q,\mu)\!\!
\int\!\! \df s\, J_\tau(s,\mu)\, \hat S_{\!\widetilde C}\,\bigg(\frac{Q C}{6}- \frac{s}{Q},\mu\bigg)\,.
\end{align}
It describes the most singular (and numerically dominating) partonic contributions in the dijet limit.
The resummation of large logarithms is achieved by evolving the functions $H_{\!Q}$, $J_\tau$, and
$\hat S_{\!\widetilde C}$ from their respective natural scales $\mu_H$,
$\mu_J$ and $\mu_S$, where logs are small, to a common scale $\mu$ (which without loss of generality can be chosen
to be, for instance, $\mu_J$). In Eq.~(\ref{eq:fact-singular}) $\hat S_{\!\widetilde C}$  is also in the $\overline{\rm MS}$ scheme, and suffers from an $\mathcal{O}(\Lambda_{\rm QCD})$ renormalon. We can switch to the renormalon-free
Rgap scheme~\cite{Hoang:2008fs} by performing subtractions on the partonic soft function (through an exponential of a
derivative operator) and simultaneously allow the same terms to change $\Omega_1^C$ from the $\overline{\rm MS}$ scheme to the Rgap scheme. Our strong coupling $\alpha_s$ will always be in the $\overline{\rm MS}$ scheme.  Adding these subtractions plus the renormalization group evolution kernels gives:
\begin{align}\label{eq:resummation}
&\frac{1}{\sigma_0}\frac{\df \hat\sigma_{\!\rm s}}{\df C} =
\frac{Q}{6} H_{\!Q}(Q,\mu_H)U_{\!H}(Q,\mu_H,\mu)\!\! \int\!\! \df s\, \df s^\prime \df k\\\nonumber
&\times\!\,
J_\tau(s,\mu_J)U_{\!J}^\tau(s-s^\prime,\mu,\mu_J)U_{\!S}^\tau(k,\mu,\mu_S) \\\nonumber
&\times
e^{-\,3\pi\frac{\delta(R,\mu_s)}{Q}\frac{\partial}{\partial C}}
\hat S_{\!\widetilde C}\,\bigg(\frac{Q C-3\pi\,\bar\Delta(R,\mu_S)}{6}- \frac{s}{Q}-k,\mu_S\!\bigg)\,,
\end{align}
where $\delta(R,\mu_S)$ is a series in powers of $\alpha_s(\mu_S)$ that can be computed directly from the thrust partonic
soft function in Fourier space. For the renormalon to be properly canceled by the subtractions, it is crucial that
the exponential and the partonic soft function are consistently expanded out to a given order in $\alpha_s(\mu_S)$. The subtractions introduce a scale $R$, which is close to the soft scale $\mu_S$ and can be used to sum up large logs in the subtraction series through the finite shift parameter $\bar\Delta(R,\mu_S)$. The  dependence on $R$ formally cancels between $\delta(R,\mu_S)$ and
$\bar\Delta(R,\mu_S)$ order by order in perturbation theory, but the $R$ parameter is crucial to eliminate the $\Lambda_{\rm QCD}$ renormalon. Nonperturbative corrections are incorporated
though a convolution with the shape function $F_{\!C}(p)$ whose first moment is $\Omega_1^C$. The hadron level prediction for the distribution is
\begin{align}\label{eq:fact-th}
\frac{1}{\sigma_0}\frac{\df \sigma}{\df C} &= \!\!\int \!\!\df p \,\frac{1}{\sigma_0}
\frac{\df \hat\sigma}{\df C}\Big(C-\frac{p}{Q}\Big)\,F_{\!C}(p)\,,\\
\frac{\df \hat\sigma}{\df C} & \,=\, \frac{\df {\hat\sigma}_{\!\rm s}}{\df C} \,+\,
\frac{\df {\hat\sigma}_{\!\rm ns}}{\df C}\,,\nn
\end{align}
and also includes the nonsingular contributions, $\df\hat\sigma_{\rm ns}/\df C$, which in the dijet limit contains all terms which are kinematically suppressed by additional powers of $C$.

For our analysis we include perturbative corrections to the matrix elements $H_{\!Q}$, $J_\tau$ and $\hat S_{\!\widetilde C}$ to $\mathcal{O}(\alpha_s^3)$. For $H_{\!Q}$ they are known analytically, whereas for $J_\tau$ and $\hat S_{\!\widetilde C}$
only the logarithmic terms at $\mathcal{O}(\alpha_s^3)$ are known (since the anomalous
dimensions are known at three loops). These non-logarithmic terms are added as unknown
coefficients that are varied when we estimate the theory uncertainties. At ${\cal O}(\alpha_s^2)$ the jet function can be directly taken
from Ref.~\cite{Becher:2006qw}. The soft function needs to be computed to $\mathcal{O}(\alpha_s^2)$~\cite{Hoang:2015hka}, which can be done analytically at $\mathcal{O}(\alpha_s)$ and for the logarithmic corrections at ${\cal O}(\alpha_s^2)$. For the non-logarithmic ${\cal O}(\alpha_s^2)$ terms our evaluation uses numerical output of the parton level MC EVENT-2~\cite{Catani:1996jh, Catani:1996vz}.

Through RGE evolution we achieve N$^3$LL resummation of the logarithmic terms. The anomalous dimensions required for solving the running equations can be taken directly from thrust. The only missing term is the four-loop cusp anomalous dimension, which is estimated using Pad\'e approximants but is nevertheless varied in a wide range when estimating perturbative uncertainties. Its effect is in any case negligibly small. The required components for a given resummation order are specified in Table~\ref{tab:orders}. We introduce a primed counting, which is defined as the regular (unprimed) one, but with the matrix elements being included to one order higher. For consistency, the renormalon subtraction series are including to one order higher as well. The primed counting achieves a better description of data and allows the correct summation of logs at the level of the distribution (for an extended discussion of this the reader is referred to \cite{Almeida:2014uva}).

We include nonsingular terms at the same order as the functions $H_{\!Q}$, $J_\tau$, $\hat S_{\!\widetilde C}$. These can be obtained by subtracting the fixed-order singular cross section as described by the SCET factorization theorem from the full QCD partonic distribution. The latter can be computed analytically at $\mathcal{O}(\alpha_s)$, and determined numerically at $\mathcal{O}(\alpha_s^2)$ and $\mathcal{O}(\alpha_s^3)$ from the parton-level MC programs EVENT2 and EERAD3~\cite{GehrmannDeRidder:2009dp, Ridder:2014wza}, respectively.
For the $\mathcal{O}(\alpha_s^2)$ and $\mathcal{O}(\alpha_s^3)$ nonsingular contributions our numerical procedure entails uncertainties which are accounted in the estimate of the theoretical uncertainty.

It is sometimes customary to write the most singular terms of an event-shape cumulant cross section
in the following exponentiated form:
\begin{align}
\hat\Sigma (C) = &\, \dfrac{1}{\hat\sigma}\!\int^{\,C}_0 \!\!\df C^\prime \dfrac{\df \hat\sigma}{\df C^\prime} =
\left( 1 + \sum_{m=1}^{\infty} B_m \left( \dfrac{\alpha_s(Q)}{2 \pi} \right)^m\, \right) \nn \\
& \times\exp \left( \sum_{i=1}^{\infty} \sum_{j=1}^{i+1} G_{i j} \left( \dfrac{\alpha_s(Q)}{2 \pi} \right)^i
\ln^j \left( \dfrac{6}{C} \right)\, \right).
\end{align}
From the result for the factorization theorem in Eq.~(\ref{eq:fact-singular}) one can determine the $G_{ij}$  and $B_i$ coefficients as shown in 
Table~\ref{tab:Gij}, see~\cite{Hoang:2014wka}.
\begin{table}[h!]
\begin{tabular}{c|c}
Resummation Order&  Calculable $G_{ij}$'s and $B_i$'s \\[1pt]
\hline
\\[-10pt]
LL  &  $G_{i,\,i+1}$ \\[1pt]
NLL$'$  &  $G_{i,\,i}$ and $B_1$\\[1pt]
N${}^2$LL$'$  &  $G_{i,i-1}$ and $B_2$\\[1pt]
N${}^3$LL$'$  &  $G_{i,i-2}$ and $B_3$\\
\end{tabular}
\label{tab:Gij}
\end{table}
\begin{table}[h!]
\begin{tabular}{l|ccccccc}
 &  cusp &  non-cusp  &  matching &  $\beta$  &  ns  &   $\delta$  \\[1pt]
\hline
\\[-10pt]
LL  &  1 &  -  &  tree  &  1  &  -    &  - \\[1pt]
NLL  &  2 &  1  &  tree  &  2  &  -    &  - \\[1pt]
N${}^2$LL  &  3 &  2  &  1  &  3  &  1   &  1 \\[1pt]
N${}^3$LL  &  4${}^\text{p}$ &  3  &  2  &  4  &  2  &  2 \\[1pt]
\hline
\\[-10pt]
NLL$'$  &  2 &  1  &  1  &  2  &  1  &  1 \\[1pt]
N${}^2$LL$'$  &  3 &  2  &  2  &  3  &  2  &  2 \\[1pt]
N${}^3$LL$'$  &  4${}^\text{p}$ &  3  &  3  &  4  &  3  &  3 \\[1pt]
\end{tabular}
\caption{Loop corrections for primed and unprimed orders. For the anomalous dimensions of $\bar\Delta(R,\mu_S)$
one uses the same orders as for other non-cusp anomalous dimensions. The superscript ``p'' indicates that a
Pad\'e approximation is being used.}
\label{tab:orders}
\end{table}
\section{Setting the Renormalization Scales}
\label{sec:profile}
\begin{figure}[t!]
\begin{center}
\includegraphics[width=0.95\columnwidth]{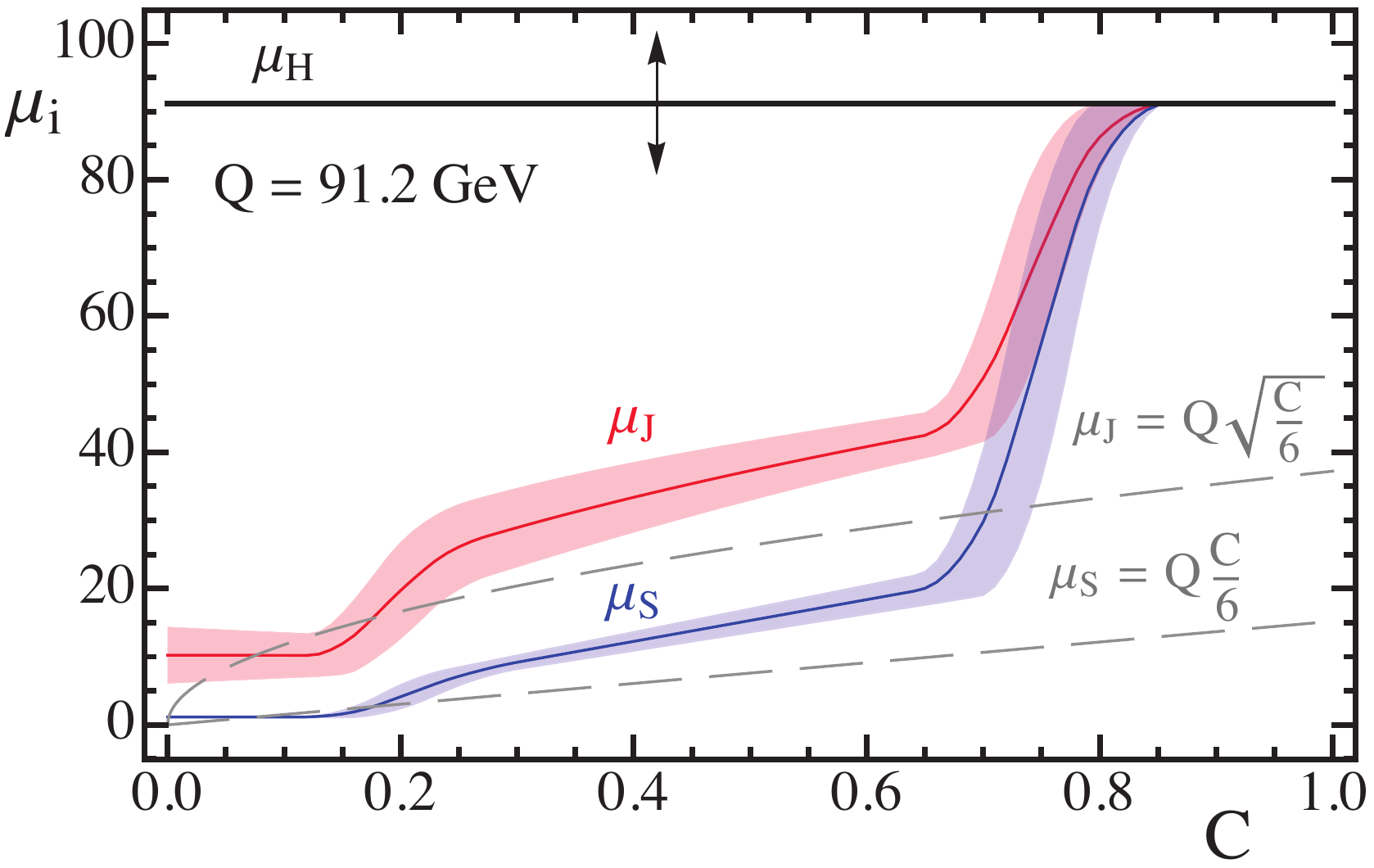}
\caption{Profile functions for the renormalization scales $\mu_H$, $\mu_J(C)$, and $\mu_S(C)$, when using the default
profile parameters (center thick line) and when varying them (light band). Fully canonical profiles are shown in gray.}
\label{fig:profile}
\end{center}
\end{figure}

The C-parameter can be divided into three distinct regions, in which the renormalization scales must satisfy different
constraints
\begin{align} \label{eq:profile-regions}
& \text{1) nonperturbative:~} C \lesssim 3\pi\,\Lambda_\text{QCD}\nn \\
&  \qquad  \mu_H  \sim Q,\  \mu_J \sim \sqrt{\Lambda_\text{QCD} Q},\  \mu_S \!\sim\! R \!\sim\!  \Lambda_\text{QCD} 
  \,, \nn \\[5pt]
& \text{2) resummation:~} 3\pi\,\Lambda_\text{QCD} \ll C < 0.75
 \\ 
& \qquad  \mu_H \sim Q,\  \mu_J \sim Q \sqrt{\frac{C}{6}},\   \mu_S \! \sim \! R \!\sim\! \frac{QC}{6}  \gg \Lambda_{\rm QCD}
  \,, \nn \\
& \text{3) fixed-order:~} C > 0.75
   \nn \\
 & \qquad \mu_H   = \mu_J  = \mu_S = R \sim Q\gg \Lambda_{\rm QCD}\nn
  \,.\end{align}
These three regions are sometimes referred to as the peak, tail and far-tail regions, respectively.
In order to satisfy these requirements we need to use renormalization scales that depend on the value of $C$, called \textit{profile functions}. The constraints in Eq.~(\ref{eq:profile-regions}) do not fully
specify the profile functions, but this ambiguity cancels order-by-order in perturbation theory. This allows 
variations of the profiles to be used to estimate perturbative uncertainties. The specific form of the profile functions and the variation of their parameters are given in Ref.~\cite{Hoang:2015hka}, and illustrated in
Fig.~\ref{fig:profile}.

\begin{figure}[t!]
\begin{center}
\includegraphics[width=0.95\columnwidth]{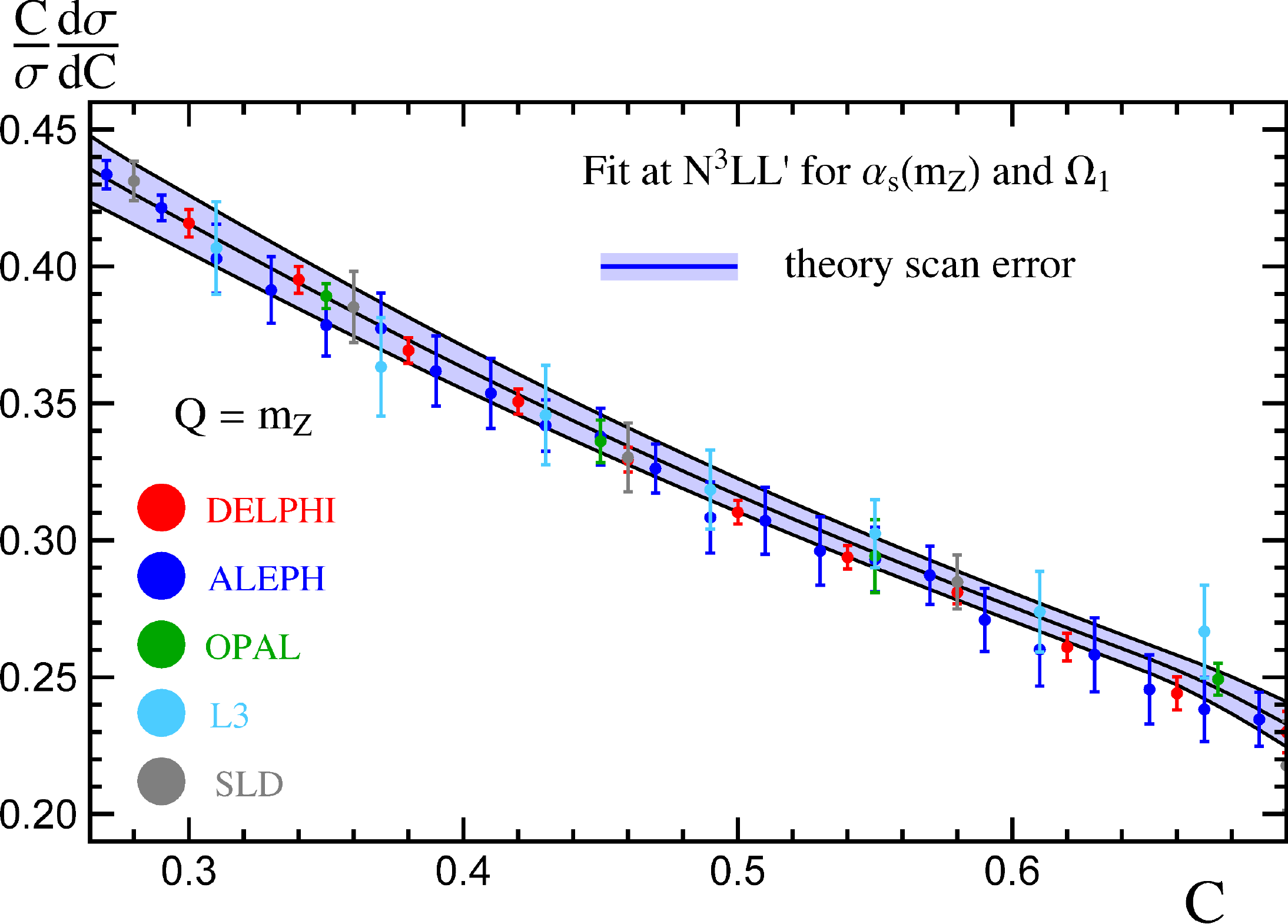}
\caption{Theoretical prediction for the C-parameter distribution at N${}^3$LL' order for $Q=m_Z$, using the best fit
values for $\alpha_s(m_Z)$ and $\Omega_1$. The blue band corresponds to the theory uncertainty as described in the text.
Experimental data for various experiments are also shown.}
\label{fig:Theory-data}
\end{center}
\end{figure}

In Fig.~\ref{fig:Theory-data} we show our C-parameter cross section for $Q=m_Z$, together with experimental data. This figure is produced with our best theoretical prediction and uses our central values for $\alpha_s(m_Z)$ and $\Omega_1$ presented in
Sec.~\ref{sec:fits}. The center blue line corresponds to the prediction for our central profiles, whereas the light blue
band shows the perturbative uncertainty.
\section{Fit results}
\label{sec:fits}
\begin{figure*}[t!]
\subfigure[]{
\includegraphics[width=0.49\textwidth]{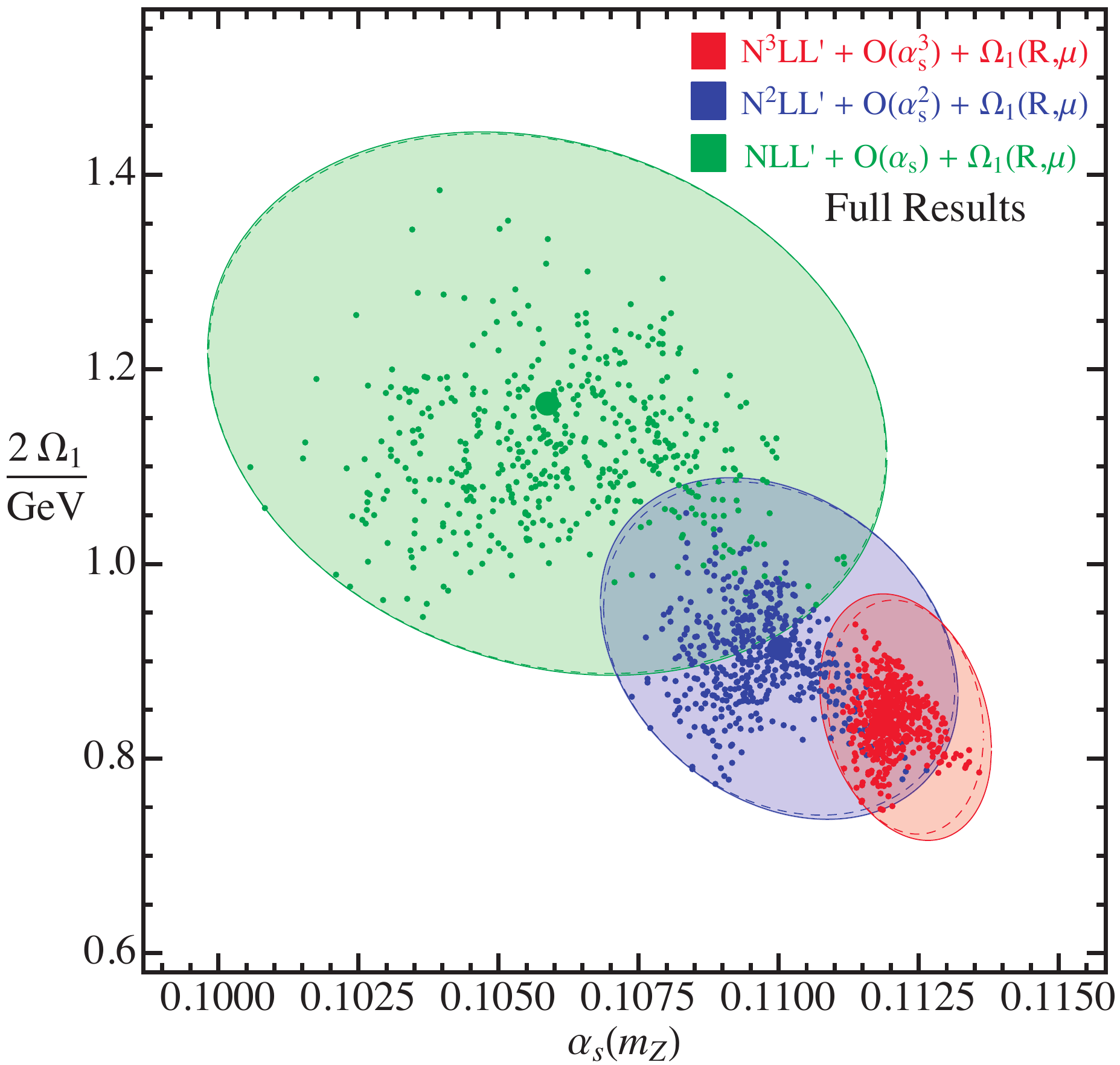}
\label{fig:alphaS}
}
\subfigure[]{
\includegraphics[width=0.48\textwidth]{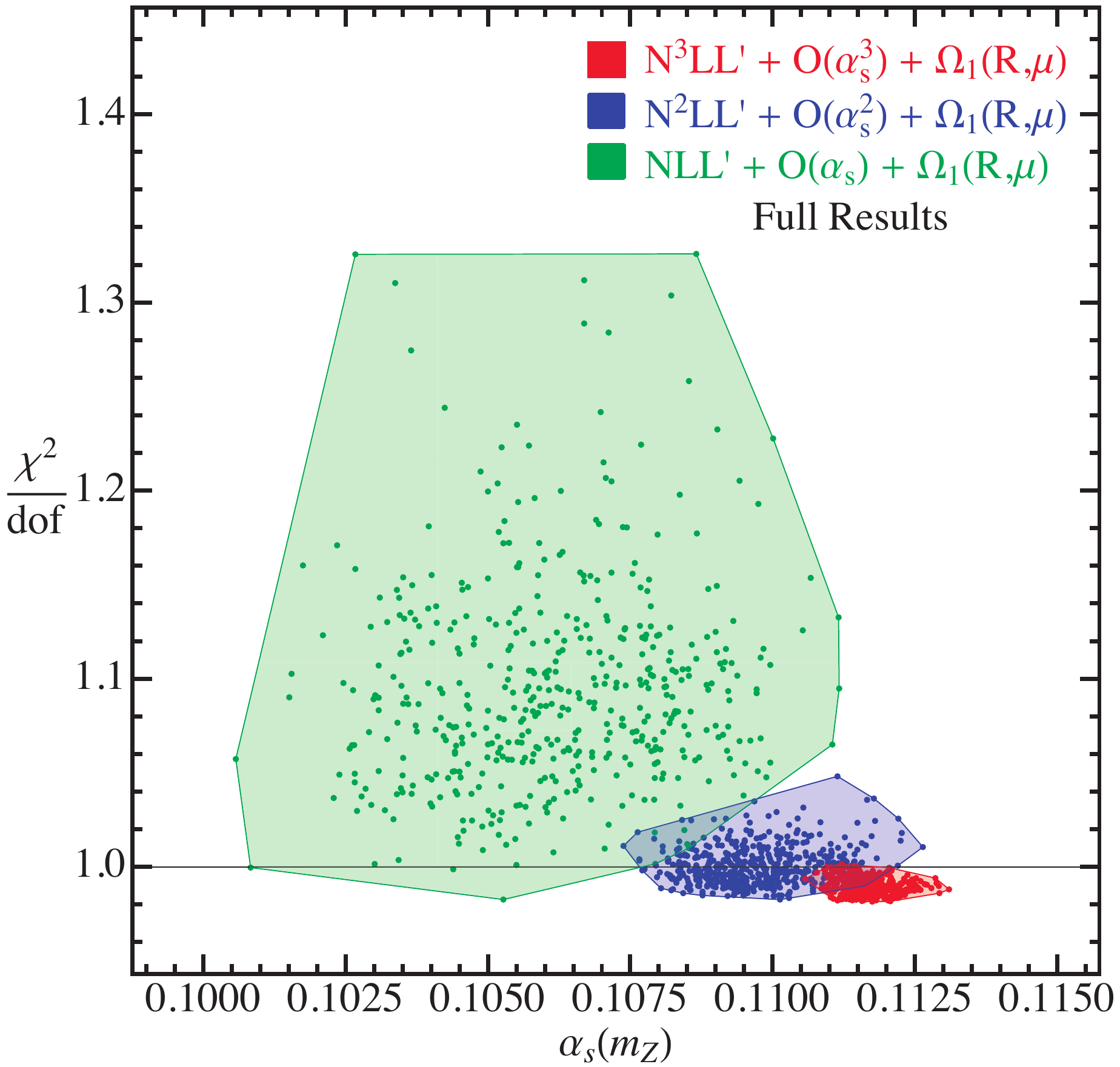}
\label{fig:chi2}
}
\vspace{-0.2cm}
\caption{The left panel (a) shows the distribution of best fit points in the $\alpha_s(m_Z)$\,-\,$2\Omega_1$ plane for fits
performed with our best theoretical predictions: resummation of large logs and power corrections defined in the
Rgap scheme with renormalon subtractions. The dashed lines corresponds to an ellipse fit to the contour of the best-fit
points to determine the theoretical uncertainty. The total (experimental\,+\,theoretical) 39\% CL standard error
ellipses are displayed (solid lines), which correspond to $1$-$\sigma$ (68\% CL) for either one-dimensional projection.
The big points represent the central values in the random scan for $\alpha_s(m_Z)$ and $2\,\Omega_1$.
The right panel (b) shows the distribution of best fit points in the $\alpha_s(m_Z)$\,-\,$\chi^2/{\rm dof}$
plane, corresponding to the points given in panel~(a).
\label{fig:alpha-chi2} }
\end{figure*}
Fitting for $\Omega_1$ together with $\alpha_s(m_Z)$ accounts for hadronization effects in a model-independent way.
In order to determine $\alpha_s(m_Z)$ and $\Omega_1$ in the same fit, one needs to perform a global analysis that includes data at many center of mass energies $Q$. For each $Q$ the differential
distribution has a noticeable degeneracy between the two fit parameters, and the use of data from the different $Q$ values breaks the degeneracy. Hence LEP and SLAC data are employed together with data from lower energy experiments such as TRISTAN and PETRA. For our analysis we use all available experimental data with energies between $35$~GeV and $207$\,GeV in the tail region.
\begin{table}[t!]
\begin{tabular}{lcc}
order &$\alpha_s(m_Z)$ (with $\overline\Omega_1$) & $\alpha_s(m_Z)$
(with $\Omega_1(R_\Delta,\mu_\Delta)$)\\[2pt]
\hline
\\[-10pt]
NLL$^\prime$             & $0.1071(60)(05)$ & $0.1059(62)(05)$ \\[2pt]
NNLL$^\prime$          & $0.1102(32)(09)$ & $0.1100(33)(06)$ \\[2pt]
N$^3$LL$^\prime$\!\!\!  & $0.1117(16)(06)$ & $\mathbf{0.1123(14)(06)}$
\end{tabular}
\caption{Best fit values for $\alpha_s(m_Z)$ at various orders with theory uncertainties from the theory scan
(first value in brackets), and experimental and hadronic error added in quadrature (second value in brackets).
Our final result at N$^3$LL$^\prime$ is shown in bold face.}
\label{tab:result}
\end{table}
To estimate theoretical uncertainties we perform $500$ fits at each order in the resummation, NLL$^\prime$, N$^2$LL$^\prime$, and N$^3$LL$^\prime$, with theory parameters randomly chosen for each fit. These parameters
specify: the profile functions, unknown perturbative coefficients, or statistical uncertainties on the numerical determination of the non-singular contributions. The result of these many fits are shown graphically as dots in Fig.~\ref{fig:alpha-chi2}. We show two projections: $\alpha_s$ vs $2\,\Omega_1$ in panel (a), and $\alpha_s$ vs $\chi_{\rm min}/{\rm dof}$
in panel (b).  As the resummation order increases the perturbative uncertainty decreases as expected, and the $\chi^2/{\rm dof}$ also decreases significantly. The corresponding numerical results and uncertainties are shown in Table~\ref{tab:result}~\cite{Hoang:2015hka}.

\begin{figure*}[tbh!]
\begin{center}
\includegraphics[width=1.5\columnwidth]{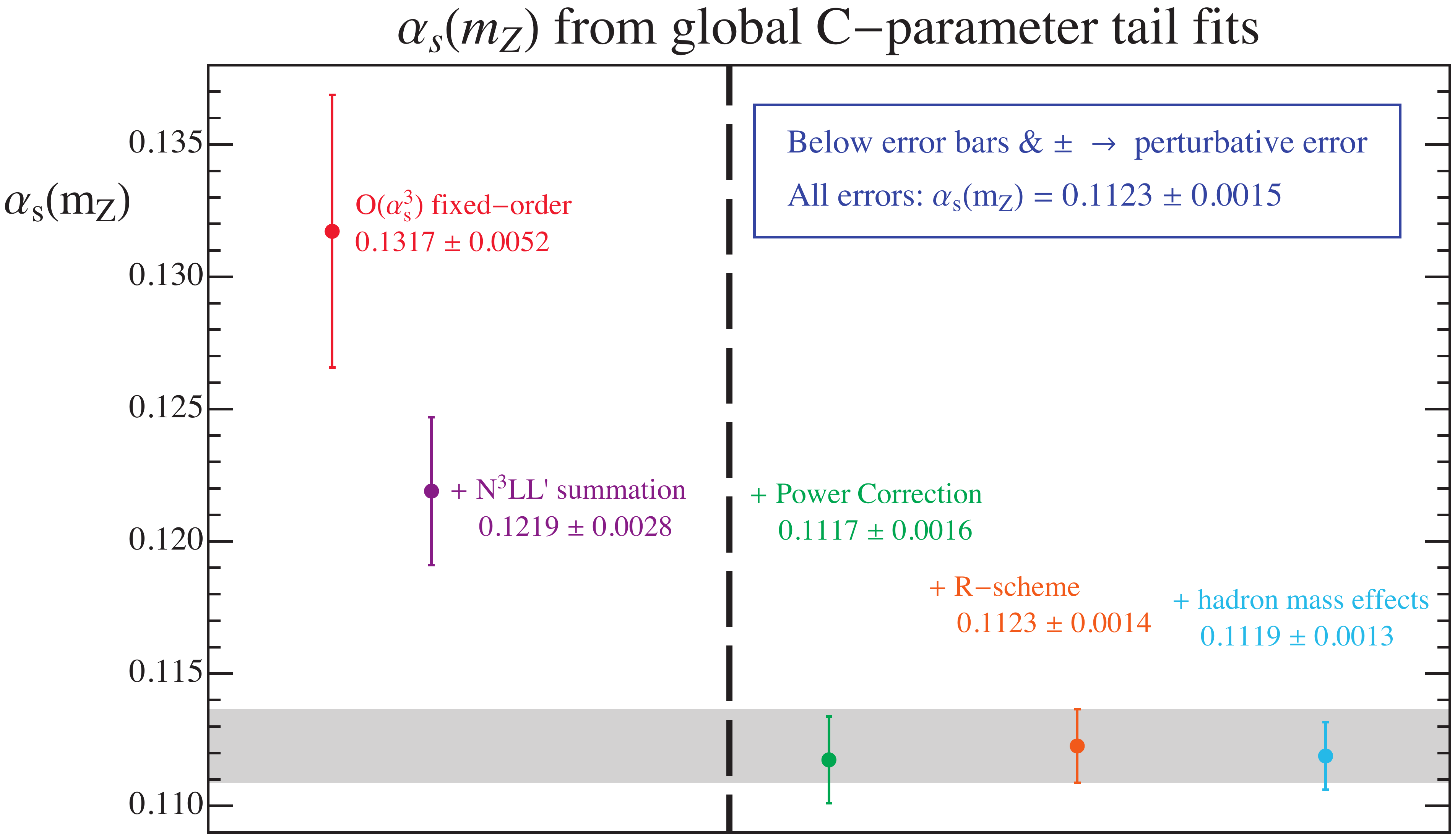}
\caption{Impact of the different components of our theoretical setup on the determination of $\alpha_s(m_Z)$.}
\label{fig:alpha-evolution}
\end{center}
\end{figure*}

In Fig.~\ref{fig:alpha-evolution} we show determinations of $\alpha_s(m_Z)$
from fits to the C-parameter distribution with different levels of theoretical sophistication. From left to right they are: fixed order with ${\cal O}(\alpha_s^3)$ (large logs not yet summed up),
N$^3$LL$^\prime$ resummation (no $\Omega_1$ in the fit), with power corrections included (not yet removing the renormalon), including Rgap scheme (not yet accounting for hadron masses), and with hadron mass effects. All error bars are perturbative, so the error bars of the first two determinations to the left of the vertical dashed line do not account for the neglect of power corrections. Including the nonperturbative power corrections by fitting $\Omega_1$ has the greatest effect on $\alpha_s(m_Z)$. Hadron mass effects give negligible contributions within current uncertainties.
\section{Conclusions}
\label{sec:conclude}
We have presented an accurate determination of $\alpha_s(m_Z)$ from fits to the C-parameter distribution. The key points to our precise theoretical prediction are: a) higher order resummation accuracy (N$^3$LL), achieved through the SCET factorization theorem, b) inclusion of $\mathcal{O}(\alpha_s^3)$ matrix elements and fixed-order kinematic power corrections, c) field-theoretical treatment of nonperturbative power corrections, and d) switching to a short-distance Rgap scheme, in which the sensitivity to infrared physics is reduced. We have not discussed hadron mass effects, as their effect is quite small. A thorough discussion can be found in \cite{Hoang:2014wka}.

Our final results from the global analysis reads\:\cite{Hoang:2015hka}
\begin{align} \label{eq:alpha-final}
\alpha_s(m_Z) & \, = \, 0.1123 \,\pm\, (0.0002)_{\rm exp}
\\ & \,\pm\, (0.0007)_{\rm hadr} \,\pm \, (0.0014)_{\rm pert}\,,
\nonumber\\  & \, = \, 0.1123 \,\pm\, (0.0015)_{\rm tot}\nn
\end{align}

\begin{figure}[tbh!]
\begin{center}
  \includegraphics[width=0.95\columnwidth]{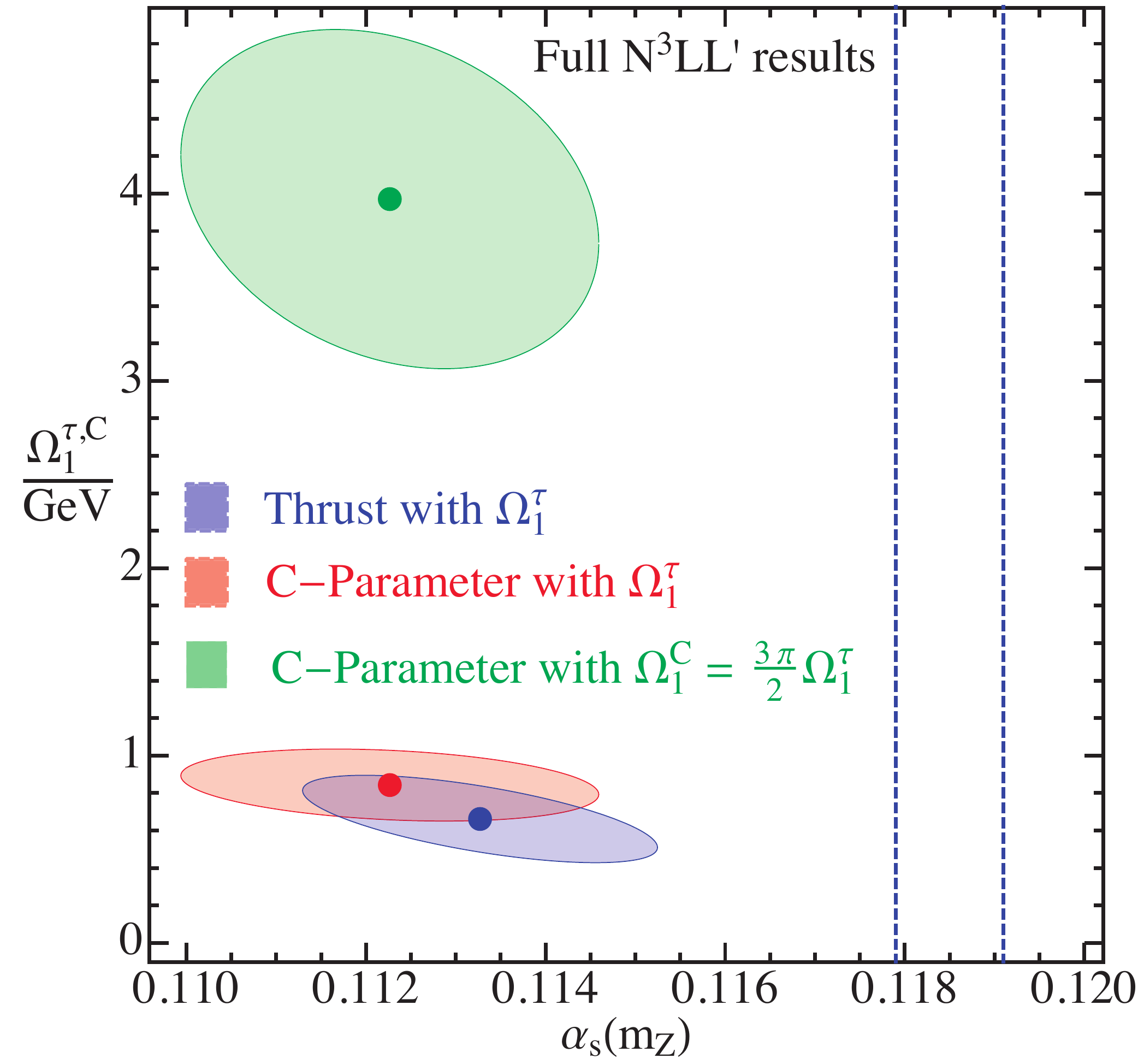}
\caption{Comparison of $\alpha_s(m_Z)$ and $\Omega_1$ determinations from fits to the C-parameter with $\Omega_1^\tau$ (red), C-parameter with $\Omega_1^C$ (green), and thrust (blue) tail cross sections, at N$^3$LL$^\prime$ with power corrections and in the Rgap scheme. The ellipses show the $\Delta \chi^2 = 2.3$ variations in the $\alpha_s(m_Z)$\,-\,$2\Omega_1$ plane, representing $1$-$\sigma$ errors for two variables.}
\label{fig:universality}
\end{center}
\end{figure}

We conclude by presenting a comparison of our \mbox{$C$-parameter} fit with the determinations of $\alpha_s$ and $\Omega_1$
from our previous thrust analysis \cite{Abbate:2010xh} of the thrust distribution, see Fig.~\ref{fig:universality}. The
figure shows that the determination of the strong coupling constant for both event shapes is compatible. Moreover there is universality in the result for the leading power correction $\Omega_1=\Omega_1^C/(3\pi)=\Omega_1^\tau/2$ determined in both analyses. The two independent determinations agree within their $1$-$\sigma$ uncertainties, where the precision of the extractions is greater than that achieved in the past. Note that without including the respective prefactors ($3\pi$ and $2$) (shown in green) the values  of $\Omega_1^\tau=0.329\, {\rm GeV}$ and $\Omega_1^C=1.98\, {\rm GeV}$ numerically differ by $\approx 4.5$-$\sigma$, so the agreement nicely demonstrates the consistency of our theoretical predictions.
A detailed comparison with other $\alpha_s(m_Z)$ determinations can be
found in Ref.~\cite{Hoang:2015hka}.

\vspace{0.5cm}
\textbf{\underline{Acknowledgments}}\\[6pt]
\indent 
We thank the Erwin-Schr{\"o}dinger Institute (ESI) for partial support in the framework
of the ESI program ``Jets and Quantum Fields for LHC and Future Colliders''.
This work was supported by the offices of Nuclear and Particle Physics of the
U.S. Department of Energy (DOE) under under Contract DE-SC0011090, and the
European Community's Marie-Curie Research Networks under contract
PITN-GA-2010-264564 (LHCphenOnet). IS was also supported in part by the Simons
Foundation Investigator grant 327942. VM was supported by a Marie
Curie Fellowship under contract PIOF-GA-2009-251174 while part of this work was
completed. AH, VM, and IS are also supported in part by MISTI global seed funds.




\nocite{*}
\bibliographystyle{elsarticle-num}
\bibliography{../thrust3}







\end{document}